\documentclass[12pt,twosides]{article}

\usepackage[spanish,english]{babel}
\usepackage{amsfonts}
\usepackage{amssymb}

\title{ {\bf Nonlinear Gravity Theories} \\ {\bf  in the} \\{\bf Metric and Palatini Formalisms}}
\author{Gonzalo J. Olmo$^{a,b}$\thanks{gonzalo.olmo@uv.es}, William Komp$^{a}$\thanks{wkomp@uwm.edu}}
\date{ {\footnotesize a) Physics Department, University of
Wisconsin-Milwaukee,}\\{\footnotesize
Milwaukee, WI 53201, USA } \\
{\footnotesize b) Departamento de F\'{\i}sica Te\'orica and
    IFIC, Centro Mixto Universidad de Valencia-CSIC.
    Facultad de F\'{\i}sica, Universidad de Valencia,}\\{\footnotesize
        Burjassot-46100, Valencia, Spain}}

\begin{document}
\thispagestyle{empty} \maketitle
\begin{abstract}
We study nonlinear gravity theories in both the metric and the
Palatini (metric-affine) formalisms.  The nonlinear character of
the gravity lagrangian in the metric formalism causes the
appearance of a scalar source of matter in Einstein's equations
that can be interpreted as a quintessence field. However, in the
Palatini case no new energy sources appear, though the equations
of motion get modified in such a way that usual matter can lead to
repulsive gravity at very low densities. Thus, the Palatini
formalism could provide a mechanism to explain the recent
acceleration of the universe without the necessity of dark energy
sources. We also show that in contrast to the metric formalism
where only the Einstein frame should be considered as physical,
the Palatini formalism allows both the Einstein and the Jordan
frames to be physically acceptable.
\end{abstract}

\newpage

\section{Introduction}

In an attempt to justify the recent observations of type Ia
supernovae \cite{Tonry03,Knop03} which indicate a late-time
accelerated expansion of the universe, certain modifications of
classical cosmology have been proposed. Some authors support the
idea that unobserved (dark) energy sources (cosmological constant,
quintessence, quantum effects, \ldots see \cite{PAD,QUINT,CDS,P-R,P-R01,
PKV} and references therein) could be responsible for this effect.
Others propose modifications of the Hilbert-Einstein lagrangian arguing
that there could be correcting terms (with an origin, perhaps, in
{\it higher} theories such as string/M-theory) that could somehow
modify Einstein's equations and justify the observations. Among
the possible modifications of the gravity lagrangian, those
consisting in nonlinear functions of the scalar curvature have
been widely studied in the literature (see \cite{WANDS} and
references therein). In particular, a correction of the type $1/R$
has been recently proposed \cite{C-T} with the purpose of
justifying this late accelerated expansion rate without the
necessity of dark energy or a cosmological constant. Such
modification has been criticised on grounds of violation of solar
system constraints \cite{CHI} imposed by the PPN formalism
\cite{WIL}. However, the possibility of using a first order
(Palatini) variational formalism that seem to avoid those
drawbacks revived the interest in this model
\cite{VOL}\cite{MENG} (see also \cite{FLA}\cite{VOL2}). \\

In some sense, the use of a nonlinear gravity lagrangian as an
alternative to dark energy sources is not completely justified. It
is well known that a metric theory of gravitation where the
lagrangian density is a nonlinear function of the scalar curvature
can be transformed (by means of a conformal transformation of the
metric) into a scalar-tensor theory, where the lagrangian is that
of Hilbert-Einstein plus a minimally coupled self-interacting
scalar field \cite{WANDS}\cite{M-S}. Thus, in this new
representation of the theory, the effects of the non-linear
lagrangian are encoded in a new source of matter, say, the scalar
field. However, a conformal transformation of the metric yields
physically inequivalent theories. This somehow forces us to
identify the physical frame before coupling matter to gravity
\cite{M-S}, i.e., one must know which metric (the original or the
transformed) represents the physical metric or if both can
represent by themselves inequivalent physical theories. The same
study should be carried out when the metric variational formalism
is abandoned in favour
of the Palatini formalism. This latter aspect is the main purpose of this paper.\\

The paper is organised as follows. First, we motivate the
discussion about the physical meaning of a frame and fix the
notation working out the subject of identifying the physical frame
within the metric formalism. We conclude that only the transformed
(Einstein) frame is physical \cite{M-S}. Then we introduce the
metric-affine formalism and show that important differences exist
between the two formalisms since in this case both the (original)
Jordan frame and the (transformed) Einstein frame can be
regarded as physical. We also discuss about the interpretation and
implications of the equations of motion when matter is introduced
in the theory. In the last sections we comment on applications to
cosmology.

\section{Frames in the Metric Formalism}

Let us consider the action of a non-linear gravity (NLG) theory in
vacuum (without matter action)
\begin{equation}\label{eq:JF0}
S[\bar{g}_{\mu \nu}]=\frac{1}{2\kappa^2}\int d^4 x\sqrt{-\bar{g}}f(\bar{R}(\bar{g}))
\end{equation}
where $\bar{g}_{\mu\nu}$ will be referred to as the Jordan frame
metric, $\bar{R}(\bar{g})\equiv \bar{g}^{\mu\nu}R_{\mu\nu}(\bar{g})$ and
$R_{\mu\nu}(\bar{g})$ is the Ricci tensor defined in terms of a connection
$\Gamma$ as
\begin{equation} \label{eq:Ricci}
R_{\mu\nu}(\bar{g})=-\partial_{\mu}
\Gamma^{\lambda}_{\lambda\nu}+\partial_{\lambda}
\Gamma^{\lambda}_{\mu\nu}+\Gamma^{\lambda}_{\mu\rho}\Gamma^{\rho}_{\nu\lambda}-\Gamma^{\lambda}_{\nu\rho}\Gamma^{\rho}_{\mu\lambda}
\end{equation}
where
\begin{equation} \label{eq:Levi-Civita}
\Gamma^{\lambda}_{\alpha\beta} =
\frac{\bar{g}^{\lambda\rho}}{2}\left[\partial_{\alpha}
\bar{g}_{\rho\beta}+\partial_{\beta}
\bar{g}_{\rho\alpha}-\partial_{\rho} \bar{g}_{\alpha\beta}\right]
\end{equation}
Varying eq.(\ref{eq:JF0}) with respect to the metric we obtain
\begin{equation}\label{eq:4-order}
f'(\bar{R}(\bar{g}))R_{\mu\nu}(\bar{g})-\frac{1}{2}f(\bar{R}(\bar{g}))\bar{g}_{\mu\nu}-
\nabla_{\mu}\nabla_{\nu}f'(\bar{R}(\bar{g}))+\bar{g}_{\mu\nu}\Box
f'(\bar{R}(\bar{g}))=0
\end{equation}
This equation can be seen as a system of fourth order differential
equations with respect to the metric. However, we can transform
eq.(\ref{eq:4-order}) into two equations where the former involves
second order derivatives of the metric and the other represents
the dynamics of a scalar field. This can be done using the so
called Helmholtz-Jordan frame
\begin{equation}\label{eq:HF0}
S[\bar{g}_{\mu \nu},\phi]=\frac{1}{2\kappa^2}\int d^4
x\sqrt{-\bar{g}}\left[f(\phi)+\left(\bar{R}(\bar{g})-\phi\right)f'(\phi)\right]
\end{equation}
where the prime of $f'(\phi)$ denotes derivative with respect to
$\phi$, and $\phi$ is a scalar field that ``on-shell'' satisfies
$\phi=\bar{R}$. This transformation of the action requires
$f''(\phi)\neq 0$. Written in this form, the action admits a
conformal transformation $g_{\mu\nu}=e^{-2\alpha}\bar{g}_{\mu\nu}$
with $e^{-2\alpha}\equiv f'(\phi)$ that leads to
\begin{equation}\label{eq:EF0}
S[g_{\mu \nu},\alpha]=\frac{1}{2\kappa^2}\int d^4x \sqrt{-g}\left[ R(g)-6\left(\nabla
\alpha\right)^2 -2V(\alpha)\right]
\end{equation}
where now\footnote{ Note that
$R(g)=e^{2\alpha}\left[\bar{R}(\bar{g})+6\bar{\Box}\alpha-6(\bar{\nabla}\alpha)^2\right]$.
See Appendix A.} $R(g)\equiv g^{\mu\nu}R_{\mu\nu}(g)$ and the
potential $V(\alpha)$ is given by the parametric equations
\begin{eqnarray}
e^{-2\alpha}&=&f'(\phi) \nonumber \\
V(\alpha)&=&\frac{\phi f'(\phi)-f(\phi)}{2(f')^2}
\label{eq:potential}
\end{eqnarray}
This frame, where the lagrangian looks like the Hilbert-Einstein
lagrangian plus a self-interacting, massless, minimally coupled
scalar field\footnote{ Note that the rescaling
$\alpha=\frac{\kappa}{\sqrt{6}}\eta$ brings the lagrangian of the
scalar field into its canonical form. We will keep the field
$\alpha$ to avoid excessive redefinitions.} $\alpha$, will be
referred to as the Einstein frame. The equations of motion in this
frame are
\begin{eqnarray}
G_{\mu\nu}(g)&=& 6\left[\nabla_{\mu}\alpha
\nabla_{\nu}\alpha-\frac{1}{2}g_{\mu\nu}\left(\nabla
\alpha\right)^2\right]-V(\alpha)g_{\mu\nu} \label{eq:metric-EF0} \\
\Box \alpha &=& \frac{1}{6}\frac{d V(\alpha)}{d \alpha}
\label{eq:alpha-EF0}
\end{eqnarray}
Since the Einstein frame metric is conformal with the Jordan frame
metric, the equations of motion in the Jordan frame (equivalent to
eq.(\ref{eq:4-order})) can be obtained from (\ref{eq:metric-EF0})
and (\ref{eq:alpha-EF0}) using the identities of Appendix A. The
results are
\begin{eqnarray}
G_{\mu\nu}(\bar{g})&=&
-2\bar{\nabla}_{\mu}\bar{\nabla}_{\nu}\alpha+4\bar{\nabla}_{\mu}\alpha\bar{\nabla}_{\nu}\alpha-\bar{g}_{\mu\nu}W(\alpha)
\label{eq:metric-JF0} \\
\bar{\Box} \alpha &=&
2\left(\bar{\nabla}\alpha\right)^2+\frac{e^{-2\alpha}}{6}\frac{d
V(\alpha)}{d \alpha} \label{eq:alpha-JF0}
\end{eqnarray}
where
\begin{equation}
W(\alpha)=\frac{f+\phi f'}{6f'}
\end{equation}

Written in this form, eqs.(\ref{eq:metric-EF0}) and
(\ref{eq:metric-JF0}) have a clear interpretation: the right hand
side represents the sources that generate the spacetime curvature
associated to their respective metrics. Though we initially
started with a pure gravity theory in vacuum, the various
transformations performed to simplify the equations of motion have
turned the vacuum gravity theory into Einstein's equations with a
scalar field as a matter source.  In \cite{M-S} the authors
consider that in the physical frame (where physical quantities are
directly measurable) this scalar degree of freedom of gravity can
be seen as a new source of matter that must satisfy the Dominant
Energy Condition (DEC) \cite{Wald}. This means that the vector
$j^{\mu}=-{T^{\mu}}_{\nu}\xi^{\nu}$ , where $\xi^{\nu}$ is a unit
vector tangent to a congruence of timelike geodesics, must be
timelike or null. This condition can be interpreted as saying that
the speed of energy flow of matter is always less than the speed
of light. It is easy to check that eq.(\ref{eq:metric-EF0})
satisfies the DEC\footnote{To get a late-time acceleration on
cosmological scales and satisfy the DEC, we require $V(\alpha)>0$.
This will be true for the NLG Lagrangian considered in section
\ref{sec:carrolnlg}}.  In fact, its right hand side coincides (up
to a constant) with the energy-momentum tensor of a
self-interacting, massless, minimally coupled scalar field and its
interpretation is straightforward. On the contrary, the right hand
side of eq.(\ref{eq:metric-JF0}) does not show the usual form of
an energy-momentum tensor. Moreover, the linear term
$\bar{\nabla}_{\mu}\bar{\nabla}_{\nu}\alpha$  can break the DEC
(it also causes difficulties in determining the ADM energy).
Consequently, the Einstein frame represents the physical frame and
matter should be coupled to gravity in this frame \cite{M-S}.

\section{Frames in the Metric-Affine Formalism \label{sec:F-MA}}

In this formalism the connection that defines the Ricci tensor is
independent of the metric. This implies that
eq.(\ref{eq:Levi-Civita}) is no longer valid and that the
conformal transformation that relates the Jordan and Einstein
metrics leaves the Ricci tensor invariant. Consequently, it turns
out that $R= e^{2\alpha} \bar{R}$ and eq.(\ref{eq:EF0}) becomes
\begin{equation}\label{eq:EF1}
S[g_{\mu \nu},\Gamma,\alpha]=\frac{1}{2\kappa^2}\int d^4x \sqrt{-g}\left[
R(\Gamma)-2V(\alpha)\right]
\end{equation}
with the potential given by eq.(\ref{eq:potential}) and $\Gamma$ is the
connection which is taken to be independent of $g$. To compute
the equations of motion for this action we have to take variations
with respect to the metric, the connection and the field $\alpha$.
The variation with respect to the connection gives
\begin{equation}\label{eq:var-gamma}
\nabla_{\beta}\left[\sqrt{-g}\left(
g^{\alpha\sigma}\delta^{\beta}_{\lambda}-\frac{1}{2}g^{\beta\sigma}\delta^{\alpha}_{\lambda}
-\frac{1}{2}g^{\alpha\sigma}\delta^{\beta}_{\lambda}\right)\right]=0
\end{equation}
Its solution \cite{MTW} implies that the Einstein frame metric
$g_{\mu\nu}$ is the ``natural'' metric associated to the
connection $\Gamma$, i.e., $\nabla_{\mu}g_{\alpha\beta}=0$ and,
therefore, $\Gamma$ is the Levi-Civita connection of the metric
$g_{\mu\nu}$ (see eq.(\ref{eq:Levi-Civita}) and replace $\bar{g}$
by $g$).

It is worth noting that eq.(\ref{eq:var-gamma}) implies
that the action (\ref{eq:EF1}) can be written as a functional of
the metric $g_{\mu\nu}$ only, i.e., that the action
\begin{equation}\label{eq:EF2}
S[g_{\mu \nu},\alpha]=\frac{1}{2\kappa^2}\int d^4x \sqrt{-g}\left[
R(g)-2V(\alpha)\right]
\end{equation}
is dynamically equivalent to eq.(\ref{eq:EF1}). Variation of
eq.(\ref{eq:EF1}) (or eq.(\ref{eq:EF2})) with respect to the
metric and the field $\alpha$ gives
\begin{eqnarray}
G_{\mu\nu}(g)&=&-V(\alpha)g_{\mu\nu} \label{eq:metric-EF1} \\
\frac{dV}{d\alpha}&=&0 \label{eq:alpha-EF1}
\end{eqnarray}
From eq.(\ref{eq:alpha-EF1}) we see that now $\alpha$ and
$V(\alpha)$ are constants. This implies that in the Jordan frame
(see Appendix A) the equations of motion are
\begin{eqnarray}
G_{\mu\nu}(\bar{g})&=&-V(\alpha)e^{-2\alpha}\bar{g}_{\mu\nu} \label{eq:metric-JF1} \\
\frac{dV}{d\alpha}&=&0 \label{eq:alpha-JF1}
\end{eqnarray}
and, therefore, there is no physical reason to exclude any of the
two frames in vacuum. Moreover, the right hand side of the
equations represents, in general, non-zero cosmological constants
that satisfy the DEC. Therefore, according to this energy
condition both frames should be regarded as good physical
candidates.

\section{ Metric-Affine formalism with matter}

In this section, we will study minimal coupling of matter to
non-linear gravity in both the Jordan and Einstein frames using
the Palatini formalism . In this formalism, the coupling is not a
trivial issue. Since gravity and matter are two different
theories, there is no a priori prescription on how to couple them.
In the metric formalism, there are many reasons to minimally
couple matter to gravity \cite{WIL}. Some of them are the
requirement that massive particles follow geodesics of the metric
and that non-gravitational experiments be independent of the
velocity and position (place and time) of the laboratory rest
frame (Einstein Equivalence Principle). In this way gravity can be
regarded as a property of curved space-time rather than as an
interaction. In the metric-affine formalism one could couple
matter to the metric, as well as to the connection, in different
ways.  We will restrict ourselves to matter lagrangians where no
explicit dependence on the connection appear. The type of coupling
to the metric will be specified in each case. We will also
indicate which variables are regarded as physical.

\subsection{ Minimal coupling in the Jordan frame}
Let us assume a minimal coupling between matter and gravity in the
Jordan frame. The corresponding action is
\begin{equation}\label{eq:MJF0}
S[\bar{g}_{\mu \nu},\Gamma,\psi]=\frac{1}{2\kappa^2}\int d^4
x\sqrt{-\bar{g}}f(\bar{R}(\Gamma))+ S_m(\bar{g}_{\mu\nu},\psi)
\end{equation}
where $\bar{g}_{\mu\nu}$ represents the physical metric and $\psi$
a set of matter fields. In the Einstein frame the above action is
equivalent to a non-minimal coupling of matter to the Einstein
metric
\begin{equation}\label{eq:NMEF0}
S[g_{\mu \nu},\Gamma,\alpha,\psi]=\frac{1}{2\kappa^2}\int d^4x \sqrt{-g}\left[
R(\Gamma)-2V(\alpha)\right]+S_m(e^{2\alpha}g_{\mu\nu},\psi)
\end{equation}
For the $\Gamma$ equations of motion, we again get eq.(\ref{eq:var-gamma}).
The equations of motion for the metric ($g$) in the Einstein frame
are given by
\begin{eqnarray}
G_{\mu\nu}(g)&=&-V(\alpha)g_{\mu\nu}+e^{2\alpha}\kappa^2\bar{T}_{\mu\nu}
\label{eq:metric-EF2} \\
\frac{d V}{d\alpha}&=&e^{4\alpha}\kappa^2\bar{T}
\label{eq:alpha-EF2}
\end{eqnarray}
where $\bar{T}_{\mu\nu}=-(2/\sqrt{-\bar{g}})\delta S_m/\delta
\bar{g}^{\mu\nu}$ is the Jordan frame energy-momentum tensor and
$\bar{T}=\bar{g}^{\mu\nu}\bar{T}_{\mu\nu}$. Using the formulas of
Appendix A, the equations of motion in the Jordan frame are
\begin{eqnarray}
G_{\mu\nu}(\bar{g})&=&-\Theta(\alpha)_{\mu\nu}-e^{-2\alpha}V(\alpha)\bar{g}_{\mu\nu}+e^{2\alpha}\kappa^2\bar{T}_{\mu\nu}
\label{eq:metric-JF2} \\
\frac{d V}{d \alpha}&=&e^{4\alpha}\kappa^2\bar{T}
\label{eq:alpha-JF2}
\end{eqnarray}
Let us interpret the two pair of equations of above. The Einstein
frame has been useful to obtain the equations of motion. The
Jordan frame is the frame where physical magnitudes appear. This
is why we have represented the matter terms using the Jordan frame
energy-momentum tensor. On the other hand, the absence of kinetic
terms for $\alpha$ in  eq.(\ref{eq:alpha-JF2}) implies that
$\alpha$ is a non-dynamical field and eq.(\ref{eq:alpha-JF2}) an
algebraic equation for $\alpha=\alpha(\bar{T})$. As a consequence,
the only matter source in eq.(\ref{eq:metric-JF2}) is that
represented by $\bar{T}_{\mu\nu}$. This is a relevant qualitative
difference with respect to the metric formalism, where the
dynamical character of $\alpha$ forced us to interpret it as an
independent source of matter that should satisfy the DEC. The
remarkable feature of eq.(\ref{eq:metric-JF2}) is that the right
hand side is not simply the energy-momentum tensor times a
constant but a more involved expression. The non-linear character
of the gravitational lagrangian manifests in the $\alpha$-terms.
If, for instance, we take the linear lagrangian\footnote{ This
example is valid even though the Helmholtz-Jordan frame needed
$f''\neq 0$.} $f(R)=R-2\Lambda$, it turns out that
$\alpha=-\frac{\ln f'}{2}=0$ and $V(\alpha)=\Lambda$ and we
recover Einstein's equations. For other lagrangians
eqs.(\ref{eq:metric-JF2}) and (\ref{eq:alpha-JF2}) represent a
natural generalisation of Einstein's equations with laboratory
matter as the only source of energy-momentum. Since for typical
sources of energy-momentum, $\bar{T}_{\mu\nu}$ (not to be confused
with the right hand side of eq.(\ref{eq:metric-JF2})) satisfies
the DEC, there is no reason
to doubt the physical character of the Jordan frame. \\

We address now the issue of conservation laws and geodesic motion
of free falling particles. It is easy to see that diffeomorphism
invariance of the action eq.(\ref{eq:MJF0}) implies that
$\bar{\nabla}^{\mu}\bar{T}_{\mu\nu}=0$. This equation guarantees
that a self-gravitating dust
$\bar{T}_{\mu\nu}=\bar{\rho}\bar{u}_{\mu}\bar{u}_{\nu}$ moves
along timelike geodesics,
$\bar{u}^{\nu}\bar{\nabla}_{\nu}\bar{u}^{\mu}=0$. If we denote by
$\bar{\tau}_{\mu\nu}$ the right hand side of
eq.(\ref{eq:metric-JF2}), one can also check that
$\bar{\nabla}^{\mu}\bar{\tau}_{\mu\nu}=0$ as
$\bar{\nabla}^{\mu}G_{\mu\nu}(\bar{g})=0$ requires. To show this
one must use that
\begin{equation}
R_{\mu\nu}(\bar{g})=e^{2\alpha}\kappa^2\left(\bar{T}_{\mu\nu}-\frac{1}{2}\bar{g}_{\mu\nu}\bar{T}\right)+
\bar{g}_{\mu\nu}\left(2(\bar{\nabla}\alpha)^2-\bar{\Box}\alpha+e^{-2\alpha}V\right)-2\left(\bar{\nabla}_{\mu}\bar{\nabla}_{\nu}\alpha
+\bar{\nabla}_{\mu}\alpha\bar{\nabla}_{\nu}\alpha\right)
\end{equation}
The contraction $R_{\mu\nu}(\bar{g})\xi^{\mu}\xi^{\nu}$, with the
unit vector $\xi^{\nu}$ tangent to timelike geodesics, leads
to\footnote{The missing terms represent total derivatives of a
scalar function along a geodesic and, therefore, vanish.}
\begin{equation}\label{eq:SEC0}
R_{\mu\nu}(\bar{g})\xi^{\mu}\xi^{\nu}=e^{2\alpha}\kappa^2\left(\bar{T}_{\mu\nu}\xi^{\mu}\xi^{\nu}+\frac{\bar{T}}{2}\right)-
e^{-2\alpha}V(\alpha)-\left(2(\bar{\nabla}\alpha)^2-\bar{\Box}\alpha\right)
\end{equation}
If, for instance, we take the Hilbert-Einstein lagrangian
($\alpha=0$,$V(\alpha)=0$) this equation becomes
\begin{equation}\label{eq:SEC1}
R_{\mu\nu}(\bar{g})\xi^{\mu}\xi^{\nu}=\kappa^2\left(\bar{T}_{\mu\nu}\xi^{\mu}\xi^{\nu}+\frac{\bar{T}}{2}\right)
\end{equation}
The condition $R_{\mu\nu}(\bar{g})\xi^{\mu}\xi^{\nu}\geq 0$, known
as {\it strong energy condition} \cite{Wald}, implies that matter
always produces attractive gravity. The addition of a (positive)
cosmological constant in the lagrangian turns eq.(\ref{eq:SEC0})
into
\begin{equation}\label{eq:SEC2}
R_{\mu\nu}(\bar{g})\xi^{\mu}\xi^{\nu}=\kappa^2\left(\bar{T}_{\mu\nu}\xi^{\mu}\xi^{\nu}+\frac{\bar{T}}{2}\right)-\Lambda
\end{equation}
In this case, the cosmological term leads to repulsive gravity
when the matter terms become negligible. By the same token,
eq.(\ref{eq:SEC0}) implies that, in general, there can be
lagrangians that lead to repulsive gravity\footnote{ Note that at
low matter densities the dominant term is the potential
$V(\alpha)$ which in vacuum becomes a cosmological constant.} .
Therefore, the metric-affine formalism could justify the recent
cosmic acceleration as a manifestation of the non-linear character
of the gravity lagrangian rather than due to the existence of {\it
dark} energy sources.

\subsection{ Minimal coupling in the Einstein frame}

If matter is minimally coupled to gravity in the Einstein frame,
the resulting action is
\begin{equation}\label{eq:MEF0}
S[g_{\mu \nu},\Gamma,\alpha,\psi]=\frac{1}{2\kappa^2}\int d^4x \sqrt{-g}\left[
R(\Gamma)-2V(\alpha)\right]+S_m(g_{\mu\nu},\psi)
\end{equation}
The variation with respect to the connection simply states the
compatibility between $\Gamma$ and the Einstein frame metric
$g_{\mu\nu}$ . Thus, this case is equivalent to the well known
Hilbert-Einstein lagrangian plus cosmological constant (see
eq.(\ref{eq:EF2}))
\begin{equation}\label{eq:MEF1}
S[g_{\mu \nu},\alpha,\psi]=\frac{1}{2\kappa^2}\int d^4x \sqrt{-g}\left[
R(g)-2V(\alpha)\right]+S_m(g_{\mu\nu},\psi)
\end{equation}
since now $\alpha$ is coupled to nothing and, therefore, $\alpha$
and $V(\alpha)$ are constants. In this case no new effects appear
and the motivation to introduce a non-linear gravity lagrangian is
lost. Obviously, this possibility is physical.

\subsection{ Non-Minimal coupling in Einstein
frame}\label{sec:NM-EF}

A glance at eq.(\ref{eq:NMEF0}) indicates that the Einstein frame
can be used to study new gravity phenomena if a non-minimal
coupling is assumed\footnote{ Superstring theory also introduces
additional fields that break the minimal coupling between matter
and gravity.}. We have shown that minimal coupling in the Einstein
frame breaks the motivation to introduce a non-linear lagrangian,
since the only result is the appearance of a cosmological
constant, which, on the other hand, can be obtained by means of a
linear function of the scalar curvature. However, if we take
eq.(\ref{eq:NMEF0}) seriously and consider the Einstein frame as
physical, i.e., with\footnote{\label{foot}Where
$T_{\mu\nu}=-(2/\sqrt{-g})\delta S_m/\delta
g^{\mu\nu}=e^{2\alpha}\bar{T}_{\mu\nu}$. In a perfect fluid, for
instance, $(P,\rho,u_{\mu})$ are the physical pressure, matter
density and four velocity in the Einstein frame, whereas
$(\bar{P}=e^{-4\alpha}P,\bar{\rho}=e^{-4\alpha}\rho,\bar{u}_{\mu}=e^{\alpha}u_{\mu})$
represent those quantities in the (now unphysical) Jordan frame.}
$g_{\mu\nu}$ and $T_{\mu\nu}$ as the physical metric and
energy-momentum tensor respectively, it turns out that the
equations of motion become
\begin{eqnarray}\label{eq:metric-NMEF1}
G_{\mu\nu}(g)&=& -V(\alpha)g_{\mu\nu}+\kappa^2 T_{\mu\nu} \\
\frac{dV}{d\alpha}&=&\kappa^2 T \label{eq:alpha-NMEF1}
\end{eqnarray}
Note that though these two equations are mathematically equivalent
to eqs.(\ref{eq:metric-EF2}) and (\ref{eq:alpha-EF2}), their
physical consequences are different, since now we are regarding
$g_{\mu\nu}$ and $T_{\mu\nu}$ as the physical variables (rather
than $\bar{g}_{\mu\nu}$ and $\bar{T}_{\mu\nu}$). The fact that the
potential $V(\alpha)$ can be seen as an {\it evolving cosmological
constant} is quite attractive
from a phenomenological point of view. \\

The conservation laws no longer lead to geodesic motion (as
expected, since conformal transformations do not preserve this
property for massive particles)
\begin{equation}
\nabla^{\mu}T_{\mu\nu}=T\partial_{\nu}\alpha
\end{equation}
The non-minimal coupling also leads to other types of violations,
such as the space-time variation of the fundamental constants
\cite{UZAN} that we do not discuss here.

\section{ Choice of $f(R)$ }

The main motivation of introducing the non-linear theory of
gravity given by $f(R)$ was to model certain cosmological
observations, (see \cite{Tonry03,Knop03}).  We know that General
Relativity (GR) is a well tested theory for solar system and
stellar astrophysical applications.  We will choose NLG
Lagrangians that will reproduce GR on such scales but will have
significant differences at cosmological scales (or densities).
With this in mind, we consider NLG Lagrangians of the form
\begin{equation}
f(R)=R-\lambda h(R)
\end{equation}
where $\lambda$ is a small parameter with suitable dimensions that
Obviously, $h$ could also depend on additional parameters, but we
are considering only the leading term of the possible
modifications. With this definition of $f$, it follows that the
conformal factor $e^{-2\alpha}$ in the {\it weak coupling} limit
($|\lambda h'|<<1$) can be expanded as
\begin{equation}
e^{-2\alpha}=f'(\phi)=1-\lambda h'(\phi)\approx 1-2\alpha
\end{equation}
and, therefore, we see that
\begin{equation}\label{eq:exp-alpha}
\alpha\approx \lambda\frac{h'(\phi)}{2}
\end{equation}
This means that in this limit, the conformal factor will be unity
plus corrections of order $\lambda$. On the other hand, the
potential is always of order $\lambda$
\begin{equation}
V(\alpha)=\lambda\frac{h-\phi h'}{(1-\lambda h')^2}
\end{equation}
The smallness of this potential illustrates the fact that a
cosmological constant (minimal coupling in Einstein frame) can be
neglected in almost all astrophysical applications, though it seems
necessary to fit cosmological observations. The equations of
motion in the case of minimal coupling in physical Jordan frame
are also compatible with Einstein's equations in this limit
\begin{equation}
G_{\mu\nu}(\bar{g}) = \kappa^2\bar{T}_{\mu\nu} +O(\lambda)
\end{equation}
In the weak coupling limit, it is evident that the case of
non-minimal coupling in the Einstein frame ({\it evolving}
cosmological constant) satisfies the geodesic motion equation up
to corrections of order $O(\lambda)$ and, therefore, is another
possible modification of GR. These limits indicate that a suitable
NLG Lagrangian can produce deviations from GR which are only
observable in cosmological densities. Those deviations could
correspond to a cosmological constant, to an {\it evolving}
cosmological constant or to a more involved modification of
Einstein's equations.

\subsection{ Example: $f(R)=R-\frac{\mu^4}{R}$}\label{sec:carrolnlg}

This lagrangian was originally proposed within the metric
formalism in \cite{C-T}, ruled out in \cite{CHI} due to
incompatibilities with the PPN formalism, and studied in the
metric-affine formalism in \cite{VOL},\cite{MENG} and \cite{FLA}.
In \cite{VOL},\cite{MENG} the Jordan frame is (implicitly) assumed
to be the physical frame (where physical magnitudes are directly
measurable). In \cite{FLA} the choice of physical variables seems
to be inconsistent\footnote{In this paper the Einstein frame
metric is (implicitly) chosen as the physical one, though the
Jordan frame matter fields represent the physical fields. To be
more consistent one should define the physical metric and fields
in the same frame. See footnote \ref{foot} in
sec.~\ref{sec:NM-EF}} and leads to unacceptable results as pointed
out in \cite{VOL2} (see however \cite{FLA2}). In this same
reference \cite{VOL2}, the author claims that it is the Einstein
frame the only physical frame. However, this conclusion is based
on the results of \cite{M-S} which we
have shown to be not applicable to the metric-affine formalism.\\

This model provides some analytical expressions which are very
useful for our discussion. Due to isotropy and homogeneity of the
Universe on cosmological scales, we will assume a perfect fluid
energy-momentum tensor minimally coupled to the Jordan frame. We
can define an energy density $\rho_{\mu}\equiv \frac{\mu^2
c^2}{8\pi G}$ in terms of which we can construct the dimensionless
quantity\footnote{ With this definition $x$ is positive.}
$x=-\bar{T}/\rho_{\mu}=\bar{\rho}_t/\rho_{\mu}$, where
$\bar{\rho}_t$ represents the total energy density in the Jordan
frame. With these definitions, eq.(\ref{eq:alpha-JF2}) and the
definition of $\alpha$ gives
\begin{eqnarray}
e^{-2\alpha}&=&
1+\frac{4}{x^2\left(1+\sqrt{1+\frac{12}{x^2}}\right)^2} \label{eq:CONF}\\
V(\alpha) &=& \rho_{\mu} e^{4\alpha}\sqrt{e^{-2\alpha}-1}
\label{eq:POT}
\end{eqnarray}
If we desire to model the late-time acceleration of the universe, then it is
reasonable to pick the order of magnitude for the mass scale of the potential
$V(\alpha)$ (the ``source'' of the acceleration in this model)
to be that of the $\Lambda$CDM \cite{PAD} and VCDM \cite{P-R01,PKV} models.
So, let $m_{\mu}c^2=\lambda{\cdot}10^{-33}$eV where $m_{\mu}$ is the mass associated
with the potential $V(\alpha)$. This gives
$\rho_{\mu}\approx \lambda^2 {\cdot}10^{-30}g/cm^3$. This density scale
is small enough to be negligible in all astrophysical situations
save for those of cosmological densities.
 In vacuum ($x\rightarrow 0$,
$e^{-2\alpha_0}=4/3$) the potential reaches its maximum value
$V(\alpha_0)\approx \rho_{\mu}/3$. In the presence of matter
($x>>1$) $e^{-2\alpha}\approx 1$ and $V(\alpha)\approx 0$. Thus, only in
interstellar medium with around $0.1$ Hydrogen atoms per cubic
centimeter and $\rho\approx 10^{-27}-10^{-28}g/cm^3$ does the
conformal factor ($e^{-2 \alpha}$) show measureable deviations from unity.
This means that, as discussed above, this theory is compatible with GR in
astrophysical applications in the three possible couplings
described above. The only differences appear at very low matter
densities and, therefore, could affect cosmological predictions.
Which of the couplings fits better the observational data will be
subject of future study \cite{K-O}. A perturbative analysis
\cite{MENG} seems to indicate a good behaviour of the physical
Jordan frame when fit to current type-Ia supernovae data.

\section{Conclusions}

We have shown that the use of a metric-affine variational
formalism in a gravity theory where the lagrangian density depends
non-linearly on the scalar curvature modifies dramatically the
results obtained using the usual metric formalism. In the metric variational
formalism, these theories lead to the appearance of a scalar field
that is only a {\it physically acceptable} source of matter in the
Einstein frame\footnote{ This new source of matter can be
interpreted as a quintessence field.}. The equations of motion are
those of GR with this additional scalar source on the right hand
side. On the contrary, the metric-affine formalism has no such
sources of matter appearing. This fact makes impossible to
distinguish between the Einstein and the Jordan frames in the
vacuum theory. When matter is minimally coupled to the metric in
the Einstein frame, the equations obtained are simply those of GR
with a cosmological constant and the particular form of the
gravity lagrangian is (almost) irrelevant. If the coupling is non
minimal, the effect is the appearance of an {\it evolving}
cosmological constant whose evolution does depend on the
particular gravity lagrangian. The third possibility discussed
here is minimal coupling carried out in the Jordan frame. In this
case, the effect of the nonlinear lagrangian is to modify the way
matter generates the space-time curvature associated with the
metric. The right hand side of the equations are no longer the
energy-momentum tensor times a constant but a more involved tensor
completely determined by the matter content. It is a remarkable
fact that a suitable choice of lagrangian can make these theories
compatible with GR in astrophysical applications. At the same
time, we have shown that these theories can lead to repulsive
gravity at very low matter densities ($\approx 10^{-27} g/cm^3$).
All these facts suggest that this formalism should be taken seriously
as an alternative to a dark energy explanation of the observed recent
accelerated expansion of the universe.\\

Though in our discussion, we have only considered the function
$h(R)=1/R$ \cite{C-T}, it seems also reasonable to take other
functions that become important at low curvatures such as $\ln R$
which might be motivated by string/M-theory
\cite{N-O}. The appropriate choice must not only agree with cosmological
observations, but has also to be consistent with laboratory
experiments \cite{FLA}. The consideration of such lagrangians
involving $1/R$ or $\ln R$ terms, in our case, was originally
motivated by the form of the effective action obtained from
quantum field theory in curved spacetime by Parker and Raval
\cite{P-R,P-R01} as well as by Parker and Vanzella\cite{P-V},
and by the success of their theory in fitting the recent
observations on the acceleration of the universe \cite{PKV}\cite{KPV}.

\section{Acknowledgements}
G.J. Olmo wants to acknowledge the Center for Gravitation and
Cosmology of the University of Wisconsin-Milwaukee for hospitality
during the elaboration of this paper and the Generalitat
Valenciana for a fellowship and financial support. G.J. Olmo
thanks Prof. Leonard Parker for very useful comments and
insights. W. Komp thanks Profs. Leonard Parker and  John L. Friedman for
helpful discussions, and the Wisconsin Space Grant Consortium and
NSF grant PHY-0071044 for their support.

\appendix

\section{ Identities}\label{Ap:1}

\begin{eqnarray}
G_{\mu\nu}(g)&\equiv&G_{\mu\nu}(\bar{g})+\Theta(\alpha)_{\mu\nu} \label{eq:id1}\\
\Theta(\alpha)_{\mu\nu}&\equiv&
2\left(\bar{\nabla}_{\mu}\bar{\nabla}_{\nu}\alpha+\bar{\nabla}_{\mu}
\alpha
\bar{\nabla}_{\nu}\alpha\right)-\bar{g}_{\mu\nu}\left(2\bar{\Box}\alpha-\bar{\nabla}_{\lambda}\alpha\bar{\nabla}^{\lambda}\alpha\right) \label{eq:id2}\\
\nabla_{\mu}\nabla_{\nu}\alpha&=&
\bar{\nabla}_{\mu}\bar{\nabla}_{\nu}\alpha+2\bar{\nabla}_{\mu}\alpha
\bar{\nabla}_{\nu}\alpha-\bar{g}_{\mu\nu}\left(\bar{\nabla}\alpha\right)^2
\label{eq:id3}
\end{eqnarray}


\begin{thebibliography}{99}

\bibitem{Tonry03}
J.L.Tonry, et al., Ap. J 594, 1 (2003)

\bibitem{Knop03}
R.A.Knop et al., Ap. J, in press (2003); astro-ph/0309368

\bibitem{PAD}
T. Padmanabhan, {\it Phys.Rept.} 380 (2003) 235-320,
hep-th/0212290

\bibitem{QUINT}
C.Armendariz-Picon, V.Mukhanov, P.J.Steinhardt {\it Phys. Rev}
{\bf D}63, 103510 (2001)

\bibitem{CDS}
R.R.Caldwell, R.Dave, P.J.Steinhardt Phys. Rev. Lett.,80,1582 (1998)

\bibitem{P-R}
L.Parker and A.Raval, {\it Phys. Rev.} {\bf D}60, 063512 (1999)

\bibitem{P-R01}
L.Parker and A.Raval, {\it Phys.Rev.Lett.} 86, 749 (2001)

\bibitem{PKV}
 L.Parker, W.Komp, and D.A.T.Vanzella,
Ap. J 588, 663 (2003)

\bibitem{WANDS}
D.Wands, {\it Class.Quant.Grav.}{\bf 11},269 (1994)

\bibitem{C-T}
S.M.Carroll, V.Duvvuri, M.Trodden and M.S.Turner, astro-ph/0306438

\bibitem{CHI}
T.Chiba, {\it Phys.Lett.}{\bf B}576 (2003) 5-11, astro-ph/0307338

\bibitem{WIL}
Clifford M. Will, {\it Living Rev.Rel.} 4 (2001) 4, gr-qc/0103036

\bibitem{VOL}
Dann Vollick, {\it Phys.Rev.}{\bf D68}(2003)06510, astro-ph/0306630

\bibitem{MENG}
X.Meng and P.Wang, {\it Class.Quant.Grav.} 20 (2003) 4949-4962,
astro-ph/0307354

\bibitem{FLA}
E.E.Flanagan, {\it Phys.Rev.Lett.} 92 (2004)071101,
astro-ph/0308111

\bibitem{VOL2}
Dann Vollick, gr-qc/0312041

\bibitem{FLA2}
E.E.Flanagan, gr-qc/0403063

\bibitem{M-S}
G.Magnano and L.M.Sokolowski,{\it Phys.Rev.}{\bf D}50 (1994)
5039-5059, gr-qc/9312008

\bibitem{Wald}
R.M.Wald, {\it General Relativity}, University of Chicago Press,
Chicago,IL, 1984.

\bibitem{MTW}
C.W.Misner, S.Thorne, and J.A.Wheeler, {\it Gravitation },
W.H.Freeman and Co., NY, 1973.

\bibitem{UZAN}
Jean-Philippe Uzan {\it Rev.Mod.Phys.}, Vol.75, No.2, April 2003.

\bibitem{K-O}
W.Komp and G.J.Olmo, {\it Work in progress}

\bibitem{N-O}
S.Nojiri and S.D.Odintsov, {\it Phys.Lett} {\bf B}576 (2003) 5-11,
hep-th/0307071 ; S.Nojiri and S.D.Odintsov, hep-th/0308176 ;
S.Nojiri and S.D.Odintsov, {\it Phys.Rev.}{\bf D} 68 (2003) 123512
, hep-th/0307288

\bibitem{P-V}
L.Parker and D.A.T.Vanzella accepted {\it Phys. Rev.} {\bf D}
(gr-qc/0312108)

\bibitem{KPV}
W.Komp, L.Parker and D.A.T.Vanzella in preparation




\end{thebibliography}
\end{document}